# Mansour's three-body force applied to nuclear matter

Hesham Mansour; FInstP.

Physics department, Faculty of Science, Cairo University, Egypt

**Abstract:**

In the present work we are going to add a correction to the BHF potential calculation by introducing a two- body density dependent potential that acts as a three –body interaction. Adding the result of this potential to the BHF potential would give us new saturation curves which coincide with the results obtained from the experiment. We compute the energy per particle E/A versus density curves for symmetric nuclear matter, pure neutron matter and the symmetry energy using modern NN-potentials like CD-Bonn (Machleidt 2001.

**Introduction:**

One of the most important problems in theoretical nuclear physics is to understand the bulk properties of nuclear matter [1-9, 14]. One of those attempts to describe such system is the BHF approximation which considers that nucleons in nuclear matter move in a mean field arising from the interaction with all other nucleons to form a bound nuclear matter.
The main goal of nuclear matter is to find the saturation curves. From the minimum of the saturation curves we can get the equilibrium binding energy and density. Nuclear matter is known to saturate at a density $\rho_0 = 0.16$ fm$^{-3}$ and Energy per nucleon E/A=-16 MeV. We also compute the binding energy per particle E/A for pure neutron matter, and the symmetry energy which is defined as the difference in Energy per nucleon between symmetric nuclear matter (An infinite system of equal number of protons and neutrons) and pure nuclear matter (An infinite system of neutrons only). Unfortunately, the BHF approach or any of its modified versions does not reproduce nither the position nor the value of the physical minimum of the energy per particle in symmetric nuclear matter. The addition of 3-body forces would improve the results. Up to now, several kinds of three-body forces TBF have been adopted in nuclear matter calculations. One is the semi phenomenological TBF such as the Urbana TBF [10]. Another TBF model adopted in the Brueckner theory is the microscopic one [11-13] based on the meson exchange theory for the NN interactions. In the BHF calculation, the TBF contribution has been included by reducing the TBF into an equivalent effective two-body interaction [13]. In the present work to improve that we add Mansour's two-body density dependent interaction which is equivalent to adding three-body forces [15-19].

In a recent published work [20] it is noticed that the author added two two-body forces for such a 3-body correction. One of them is density independent which is not to be considered as a three-body force contrary to what the author claims. Also the free energy which was calculated according to Eq. (13) is true because it contains two variables m* and T. But, the author used it in a wrong way. It was taken at constant values for m* where m* is in fact a function of the density. So the figures (3), (7 – 9) are drawn with this wrong method. Also, the effect of temperature is not being real with this method [21, 22]. Also the calculation of symmetry energy according to equation 19 is not real especially at high densities [23] contrary to the claim of the whole work.

## Correction to the BHF potential:

In this work we introduce a density dependent Skyrme effective interaction term in addition to the BHF potential Eq.1. This is a two-body density dependent potential which acts as a three-body interaction. Where $t_i$ and $x_i$ are interaction parameters and $p_\sigma$ is the spin exchange operator, $\rho$ is the density and $r_1$, $r_2$ are the position vectors for the two particles. $\alpha_i = (1/3, 2/3, 1/2, \text{ and } 1)$. Using the potential given by Eq.1 one can calculate E/A for nuclear matter and neutron matter. Besides, one can calculate the pressure P and the symmetry energy. The expressions for E/A and P for nuclear matter depend only on $t_1$ and $t_2$. Two equations are solved simultaneously for ($\alpha_i=1/3$, and $\alpha_i=2/3$) in order to find the two parameters $t_1$ and $t_2$ Where E/A $_{new}$ at saturation density is equal to -16 MeV and pressure $P_{new}$ at saturation density is equal to 0. . The values of $t_1$ & $t_2$ equals 10.473, -6.548 for CD-Bonn potential. Using the values of $t_1$ and $t_2$ we can obtain the correction to the BHF [23] and find the new corrected energy per nucleon which is shown in the following figure.

$$V(\overline{r_1}, \overline{r_2}) = \sum_{i=1}^{4} t_i (1 + x_i P_\sigma) \rho^{\alpha_i} \delta(\overline{r_1} - \overline{r_2}) \qquad (1)$$

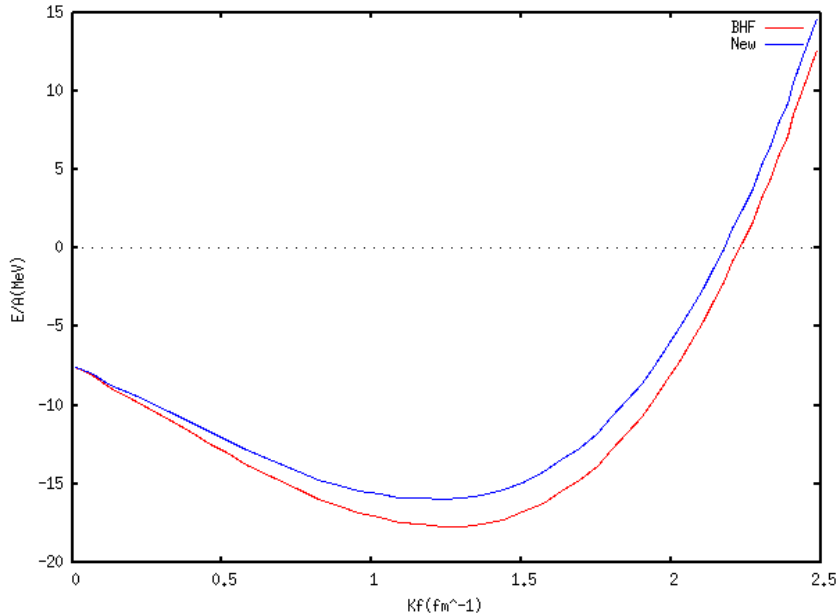

Figure (1): Energy per nucleon in MeV vs. Fermi momentum in fm$^{-1}$
for symmetric nuclear for both the BHF and the new values of Energy per nuclear for CD-Bonn potential.

As we can see the new saturation density can be adjusted to coincide with the required experimental values E/A=-16 MeV and $K_f$=1.33 fm$^{-1}$.

**Symmetry Energy:**

In order to find the values of $x_1$ and $x_2$ we take two points from the experimental symmetry energy curves and take their corresponding values in the BHF [23] thus we can have two equations where we can solve for $x_1$ and $x_2$. In case of CD-Bonn potential the experimental symmetry energy curve is shown in figure(2). The two points taken from the experimental symmetry energy curve are $\rho_1$=0.0776 fm$^{-3}$ and $\rho_2$=0.27068 fm$^{-3}$. We transform the values of the density into Fermi momentum k [fm$^{-1}$] by using the relation $\rho=2k^3/3\Pi^2$ and get $k_1$=1.047466 fm$^{-1}$, for the symmetry energy $E_1$=19.2857 MeV and $k_2$=1.588364 fm$^{-1}$, for the symmetry energy $E_2$= 45 MeV. The values of symmetry energy in BHF corresponding to the two points are 14.6992 MeV and 28.8349 MeV respectively. We calculate the correction factor from our potential and add it to the BHF symmetry energy and the result is the new symmetry energy values to be compared with the experimental curve. We find that $x_1$=18.3404 and $x_2$=34.64078. Calculating the new symmetry energy and plotting it against Fermi momentum for CD potential we get new corrected symmetry energy. As one can see from Figure 2, our calculation coincides with the experimental data.

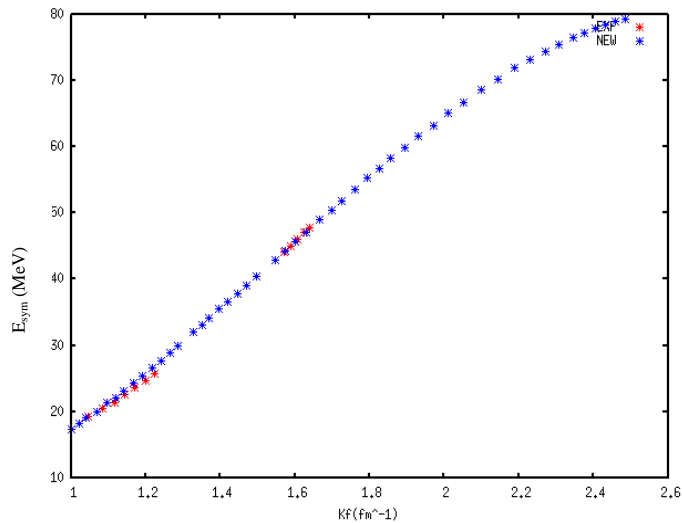

Figure (2): New Symmetry Energy (MeV) plotted against Fermi momentum k (fm$^{-1}$) along with the symmetry energy obtained from the experiment for CD- Bonn Potential. Experimental data are from reference 25

The next step is to calculate the E/A graphs for the pure neutron matter. By using the values obtained for interaction parameters $t_1, t_2, x_1,$ and $x_2$, one can calculate E/A same way as before. The results are shown in the following graph for the CD-Bonn potential.

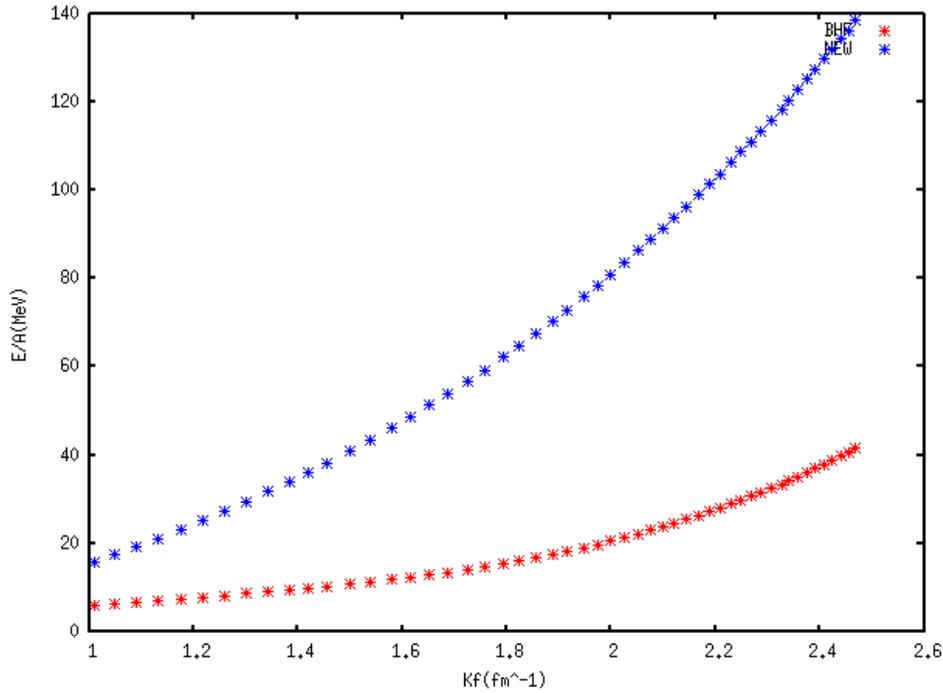

Figure (5): Energy per nucleon (MeV) versus Fermi momentum $k_f$ (fm$^{-1}$) for pure neutron matter for the BHF and the new E/A for CD-Bonn potential.

**Conclusion:**

Adding a two-body density dependent potential that acts as three-body interaction to BHF potential would produce saturation curves that coincide with the correct values of energy per nucleon and density of nuclear matter. The value of the energy per nucleon at saturation point we obtained is -16 MeV at $\rho_0 = 0.16$ fm$^{-3}$ density or Fermi momentum $K_f$=1.33 fm$^{-1}$ which is the correct value of the saturation point. It is noted that Mansour's potential given by equation 1. is resilient and very useful in reproducing all the physical quantities related to nuclear and neutron matter properties over a wide density range [24].